\documentclass[a4paper]{jpconf}
\usepackage{graphicx}
\usepackage{amsmath}
\usepackage{amssymb}
\usepackage{subfigure}

\begin{document}
\title{A Blast Wave Model With Viscous Corrections}

\author{Z Yang and R J Fries}

\address{Cyclotron Institute and Department of Physics and Astronomy, Texas A\&M University, College Station TX 77843, USA}

\ead{zdyang@physics.tamu.edu}

\begin{abstract}
Hadronic observables in the final stage of heavy ion collision can be described well by fluid dynamics or blast wave parameterizations. We improve existing blast wave models by adding shear viscous corrections to the particle distributions in the Navier-Stokes approximation. The specific shear viscosity $\eta/s$ of a hadron gas at the freeze-out temperature is a new parameter in this model. We extract the blast wave parameters with viscous corrections from experimental data which leads to constraints on the specific shear viscosity at kinetic freeze-out. Preliminary results show $\eta/s$ is rather small.
\end{abstract}

\section{Introduction}
In the initial stage of heavy-ion collisions, a hot and dense fireball is created. The fireball cools rapidly and expands into the surrounding vacuum. This process continues until it reaches a final freeze-out stage when hadrons decouple from each other. Hadronic observables in the final stage of the heavy-ion collisions can be described well by fluid dynamics or blast wave parameterizations. In this contribution, we construct a blast wave model with viscous corrections by calculating the viscous stress tensor from the parameterized flow field in the Navier-Stokes approximation, similar to the models developed in \cite{1}\cite{2}.
\section{Viscous Corrections In Fluid Dynamics And Blast Wave}
Fluid dynamics models the time evolution of the nuclear matter in heavy ion collisions, determined by initial conditions and the equation of state. When the system becomes dilute enough, kinetic freeze-out happens at a temperature $T=T_{\mathrm{fo}}$. At the kinetic freeze-out fluid cells are translated into particles through the Cooper-Frye formula
\begin{equation}
 E\frac{d^3N}{dp^3}=\frac{g}{(2\pi)^3}\int_\sigma f(p,T)p^{\mu}d\sigma_{\mu}\, .
\end{equation}
Shear viscous corrections, and in particular the specific shear viscosity $\eta/s$ will enter both the fluid dynamic equations of motion and the Cooper-Frye formula at freeze-out. Extractions of $\eta/s$ from data using viscous fluid dynamics thus conflate the effects of shear viscosity on the dynamical evolution and on freeze-out. This is to some extent also true in calculations that switch to a hadronic transport model for freeze-out. Typical values extracted for $\eta/s$ from fluid dynamics or fluid dynamics plus hadronic transport are $(1-2)/4\pi$ \cite{3}, averaged over all temperatures and including freeze-out in the averaging. Efforts are on the way to pin down the temperature evolution of $\eta/s$ during the dynamical evolution \cite{4}. In contrast, the blast wave model fits the flow field and shape of the fireball in the transverse plane. It becomes then sensitive to $\eta/s$ at kinetic freeze-out only.
\section{Viscous Blast Wave Parameterization}
The hadron spectrum at freeze-out is given by Eq.\ (1) above where $f(p,T)$ is the distribution function of the given hadron species, $g$ is its degeneracy and $\sigma^\mu$ is the normal vector on the freeze-out hypersurface. In equilibrium $f(p,T)=f_0(p,T)$ and with shear viscous corrections $f(p,T)=f_0(p,T)+\delta f$, where
\begin{equation}
f_0(p,T)=\frac{1}{e^{p\cdot u/T}\pm 1}
\end{equation}
\begin{equation}
\delta f=\frac{1}{2}\frac{p_\mu p_\nu}{T^2}\frac{\pi^{\mu \nu}}{e+p}f_0(p,T)
\end{equation}
and $\pi^{\mu \nu}$ is the shear stress tensor \cite{5}\cite{6}. In the first-order (Navier-Stokes) approximation, $\pi^{\mu \nu}$ is given by
\begin{equation}
\pi^{\mu \nu}=2\eta\langle\partial ^{\mu}u^{\nu}\rangle
\end{equation}
we plug Eq.\ (4) into Eq.\ (3) and find
\begin{equation}
\delta f=\frac{1}{T^3}\frac{\eta}{s}p_{\mu}p_{\nu}\langle\partial ^{\mu}u^{\nu}\rangle f_0(p,T)\, .
\end{equation}
We have neglected a Bose/Fermi enhancement/suppression factor since we will not work at very small momenta $p\lesssim T$. We see that the viscous correction at freeze out is directly proportional to $\eta/s$.

We choose constant $\tau=\sqrt{t^2-z^2}$ as the freeze-out hypersurface. We follow \cite{7} and parameterize the blast wave as follows. The coordinates are given by
\begin{equation}
x^{\mu}=(t,x,y,z)=(\tau \cosh\eta,\rho R_x\cos\phi,\rho R_y\sin\phi,\tau \sinh\eta)
\end{equation}
where $\rho=\sqrt{\frac{x^2}{R_x^2}+\frac{y^2}{R_y^2}}$ is the normalized radial coordinate. The hadron momentum is 
\begin{equation}
p^{\mu}=(m_T\cosh Y,p_T\cos\theta,p_T\sin\theta,m_T\sinh Y)
\end{equation}
where $m_T=\sqrt{m^2+p_T^2}$ is the transverse mass and $Y$ is momentum rapidity. We will work with hadrons at $Y=0$ here. The flow velocity can be written as \cite{5}\cite{7}
\begin{equation}
u^{\mu}=(\cosh\eta_L\cosh\eta_T,\sinh\eta_T\cos\phi_b,\sinh\eta_T\sin\phi_b,\sinh\eta_L\cosh\eta_T)
\end{equation}
where $\eta_L$ is given by  the longitudinal velocity $v_L=\tanh\eta_L$,  $\eta_T$ is given by the transverse velocity $v_T=\tanh\eta_T$. For longitudinal flow, we choose $v_L=\frac{z}{t}$, thus $\eta_L=\eta$, which enforces boost invariance. For transverse flow, we  use the parameterization
\begin{equation}
v_T=\alpha \rho^n, \qquad\alpha=\alpha_0+\alpha_2\cos2\phi_b
\end{equation}
where $\alpha$ is the surface velocity, $\alpha_0$ is the average surface velocity, $\alpha_2$ is an elliptic deformation of the flow field and $n$ is a power term. Let us summarize the parameters in this blast wave: freeze-out temperature $T$, surface velocity $\alpha_0$, velocity profile power $n$, velocity deformation $\alpha_2$, ratio of event plane radii $R_y/R_x$ and specific viscosity $\eta/s$.

We are left to calculate the Navier-Stokes shear stress tensor given our velocity field. In detail this tensor is $\pi^{\mu \nu}=2\eta\langle\partial^{\mu}u^{\nu}\rangle$ where \cite{6}
\begin{equation}
\langle\partial ^{\mu}u^{\nu}\rangle\equiv[\frac{1}{2}(\Delta_{\sigma}^{\mu} \Delta_{\nu}^{\tau})-\frac{1}{3}\Delta^{\mu \nu}\Delta_{\sigma \tau}]\partial^{\sigma}u^{\tau}, \qquad
\Delta^{\mu \nu}=g^{\mu \nu}-u^\mu u^\nu\, .
\end{equation}
The computation of the Navier-Stokes shear tensor is now straight forward but results in long expressions. We show one derivative as an example 
\begin{equation}
\partial_2u^1=\frac{\partial u^1}{\partial y}=n\sinh\eta_T\cosh^2\eta_T\frac{\sin\phi}{\rho R_y}\cos\phi_b-\sinh\eta_T\frac{\tan^2\phi_b}{(1+\tan^2\phi_b)^{3/2}}\frac{1}{\rho R_y\sin\phi}\, .
\end{equation}
\section{Extraction Of Freeze Out Shear Viscosity From Data}
To extract the blast wave parameters from experimental data, we use a Bayesian package from the Models and Data Analysis Initiative (MADAI) project \cite{4}\cite{8}. We first calculate training points in parameter space given a prior range for each parameter. The package then uses a Gaussian process emulator to estimate spectra at random parameter values. Finally it will do a likelihood analysis and give the maximum likelihood parameters.

The data used here is from the ALICE collaboration for Pb+Pb collisions at 2.76 TeV in the 40$\%$-50$\%$ centrality bin \cite{9}\cite{10}. Our fit ranges are 0.325-2.75 GeV/$c$, 0.225-1.95 GeV/$c$ and 0.425-1.25 GeV/$c$ for proton, kaon and pion respectively. We use both the transverse momentum spectra (without their absolute normalization) and elliptic flow $v_2$ to constrain the blast wave parameters. We set $\tau$=9 fm/$c$, $R_xR_y$=12 fm$^2$ and assume $\partial_\tau u^\mu$=0. 
\begin{figure}[b]
\centering
\includegraphics[height=12.6cm,width=\textwidth]{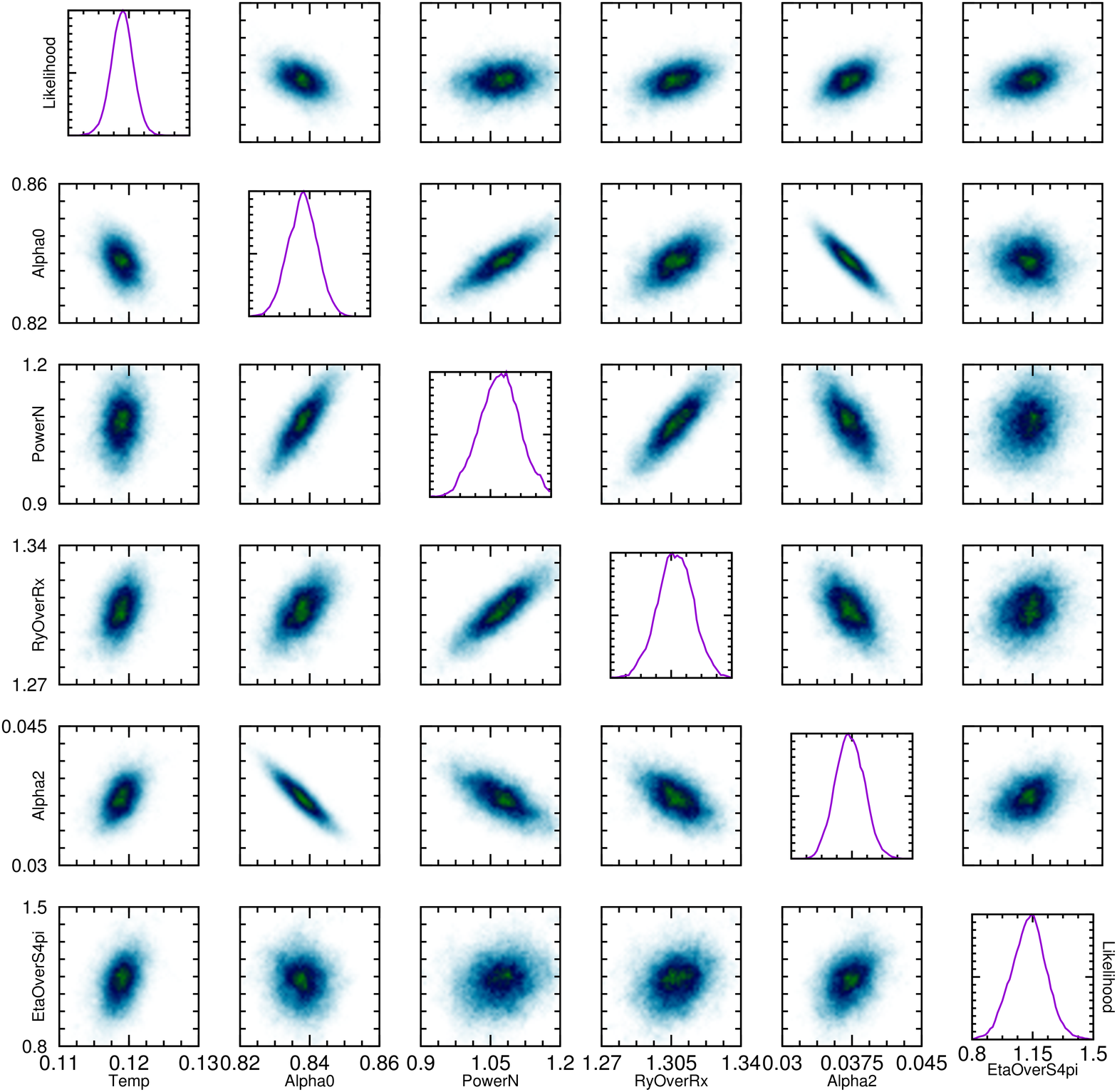}
\caption{Likelihood analysis of our fit. The x and y axes show our chosen prior ranges for all parameters. The plots on the diagonal show likelihood distributions and the off diagonal plots show correlations.}
\end{figure}

The likelihood analysis given by the MADAI package is shown in Fig.\ 1. The best  parameter values are $T$=0.119 GeV, $\alpha_0$=0.838$c$, $n$=1.07, $R_y/R_x$=1.31, $\alpha_2$=0.0372$c$, $\eta/s$=1.13/4$\pi$. We can check the quality of our fit results by calculating the transverse momentum spectra, see Fig.\ 2 and elliptic flow $v_2$, see Fig.\ 3 for the best parameter values. 

The $p_T$ spectra with the best fit parameters describe the data rather well. Elliptic flow is generally described satisfactorily as well, but there is slight tension between the kaons and pions. This can be ameliorated by allowing pions to have a separate (lower) freeze-out temperature which improves the fits (not shown here) and leads to a very similar value of the specific shear viscosity. 
\begin{figure}[t]
\centering
\includegraphics[height=1.43in,width=2 in]{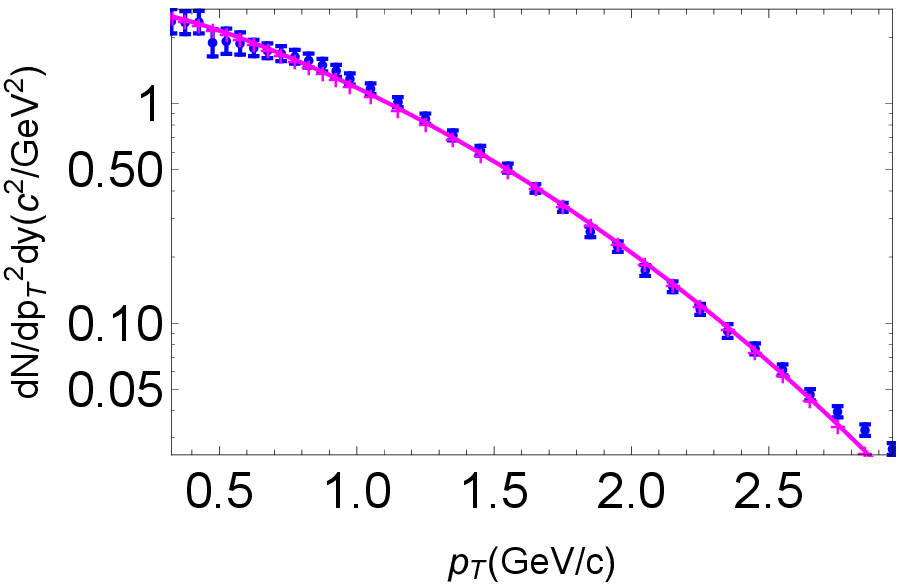}
\includegraphics[height=1.43in,width=2 in]{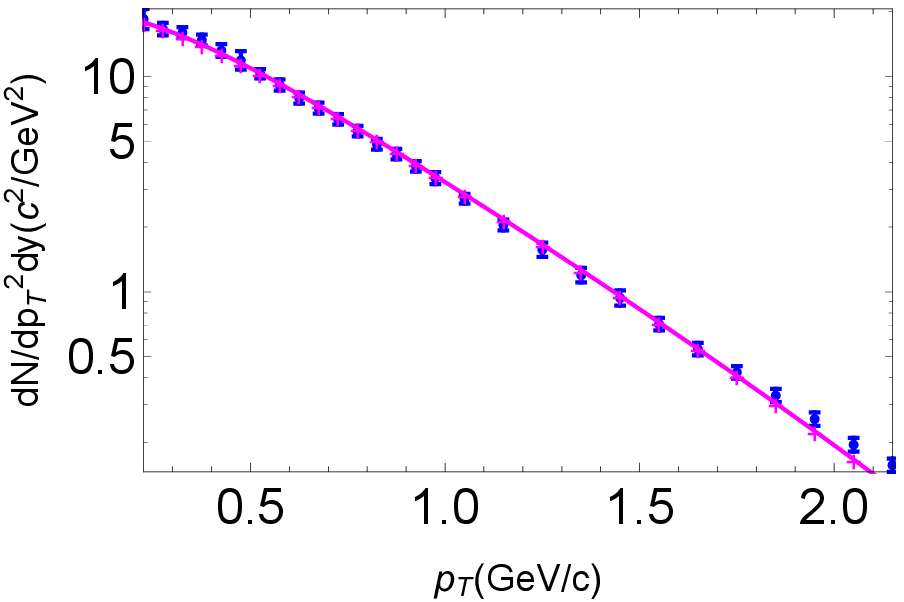}
\includegraphics[height=1.43in,width=2 in]{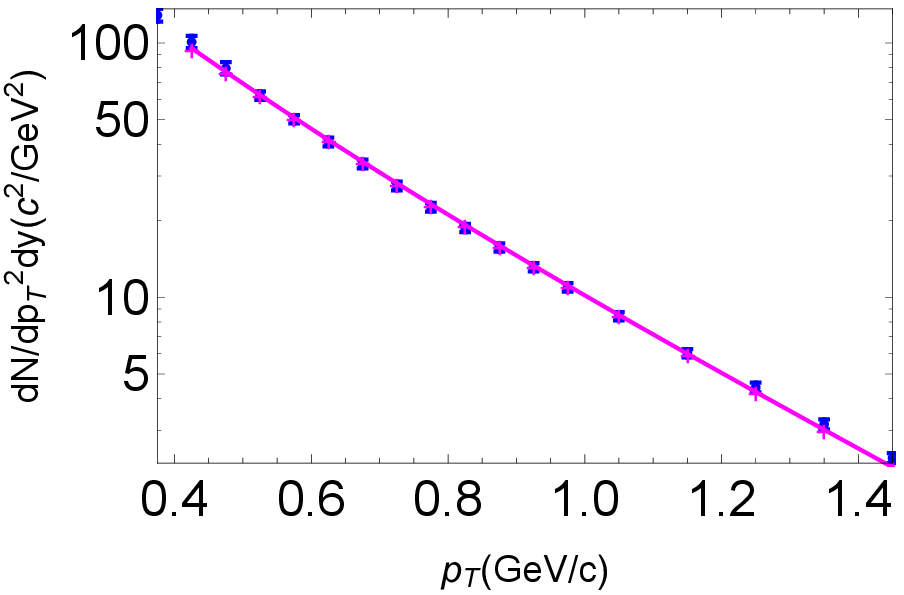}
\caption{Transverse momentum spectra for protons (left), kaons (center) and pions (right). Solid line: viscous blast wave result with best fit parameters. Data: ALICE collaboration \cite{9}.}
\end{figure}
\begin{figure}[t]
\centering
\includegraphics[height=1.43in,width=2in]{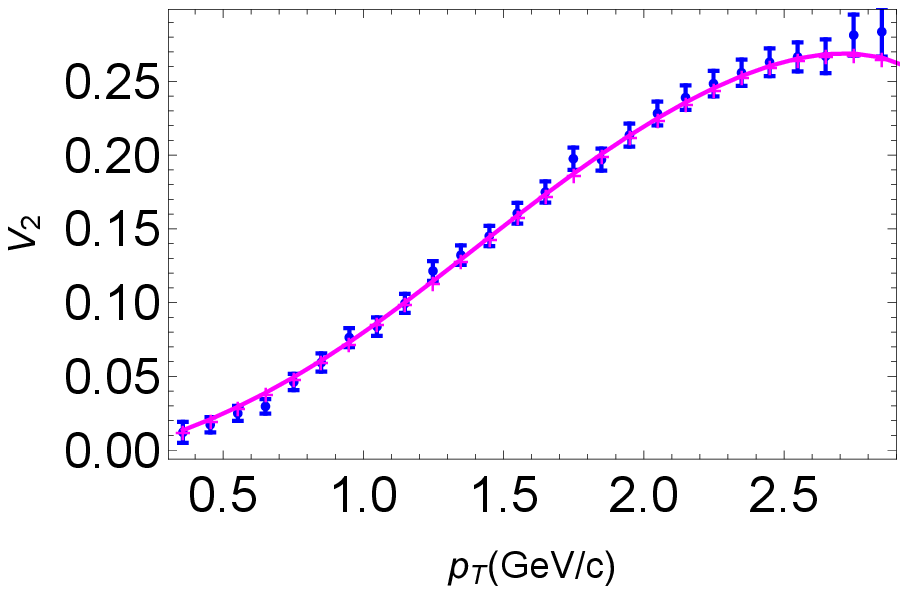}
\includegraphics[height=1.43in,width=2in]{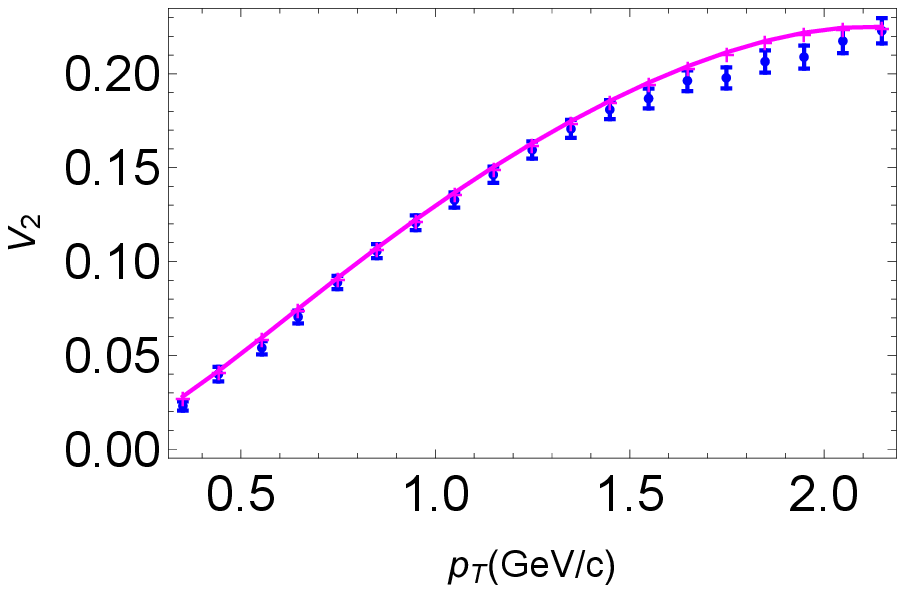}
\includegraphics[height=1.43in,width=2in]{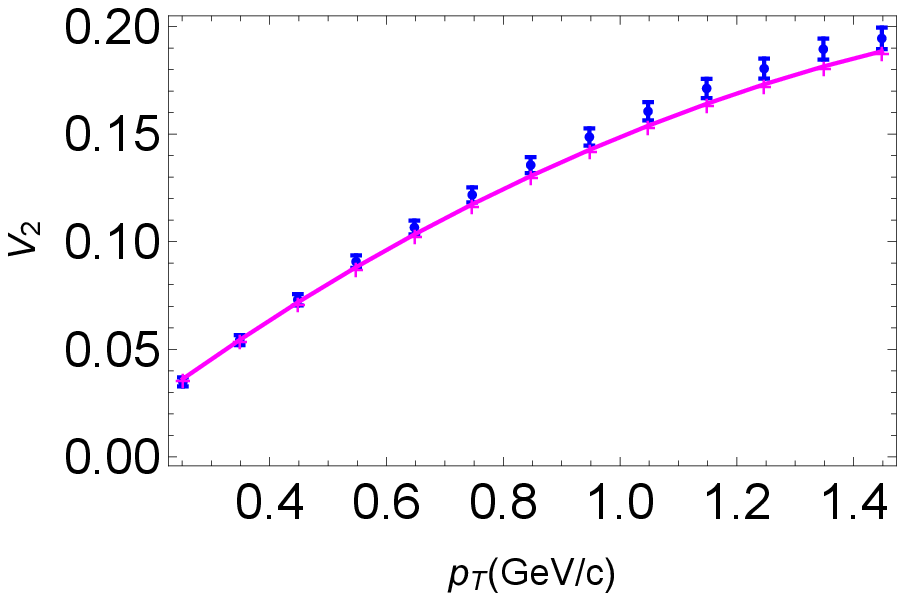}
\caption{Same as Fig.\ 2 for elliptic flow $v_2$. Data: ALICE collaboration \cite{10}.}
\end{figure}

\section{Conclusions}
We have developed a viscous blast wave model. We have used this model to extract $\eta/s$ at kinetic freeze-out in a complimentary way. The preliminary results show $\eta/s$ is rather small. In the future, we will improve our statistical analysis and work on more centrality bins. We may also use the viscous blast wave for other projects like quark recombination. This work was supported by the U.S.\ National Science Foundation.

\end{document}